# Multimodal nonlinear optical polarizing microscopy of long-range molecular order in liquid crystals

**Taewoo Lee, Rahul P. Trivedi, and Ivan I. Smalyukh[*]**

*Peer Department of Physics, Liquid Crystal Materials Research Center, and Renewable & Sustainable Energy Institute, University of Colorado, Boulder, CO 80309, USA*

[*]*Corresponding author: ivan.smalyukh@colorado.edu*

We demonstrate orientation-sensitive multimodal nonlinear optical polarizing microscopy capable of probing orientational, polar, and biaxial features of mesomorphic ordering in soft matter. This technique achieves simultaneous imaging in broadband coherent anti-Stokes Raman scattering, multi-photon excitation fluorescence, and multi-harmonic generation polarizing microscopy modes and is based on the use of a single femtosecond laser and a photonic crystal fiber as sources of the probing light. We show the viability of this technique for mapping of three dimensional patterns of molecular orientations and that images obtained in different microscopy modes are consistent with each other. © 2008 Optical Society of America

Soft matter systems such as liquid crystals (LCs) exhibit polymesomorphism of phase behavior combined with varying degrees of orientational and positional ordering intermediate between that of isotropic fluids and crystalline solids. [1] The long-range orientational order is a salient feature of these systems that results in an unprecedented richness of ground-state structures and textural behavior associated with the uniform alignment of molecules on the scale of nanometers and their slowly varying patterns on the scale of microns. In uniaxial nematic LCs, local average molecular orientations are described by the director field **n(r)**, which is also the optical axis. LC ordering can be polar or non-polar, uniaxial or biaxial, and with varying degrees of positional ordering. There are, however, no three dimensional (3D) labeling-free optical imaging techniques for probing all of these ordering features at the same time, although a substantial progress in LC imaging has been achieved by the use of fluorescence confocal polarizing microscopy (FCPM) [2] and several nonlinear optical microscopy techniques, such as second harmonic generation (SHG), [3] third harmonic generation, [4] multi-photon excitation fluorescence (MPEF), [5] sum frequency generation, [6] and coherent anti-Stokes Raman scattering polarizing microscopy (CARS-PM). [7-9] We describe a multimodal nonlinear optical polarizing microscopy (MNOPM) allowing for such non-invasive 3D imaging by combining simultaneous broadband CARS, MPEF and multi harmonic generation (MHG) imaging modalities [10,11] with the achromatic control of polarization of excitation beams using twisted nematic polarization rotator [12].

The schematic diagram of the setup is shown in Fig. 1. A tunable (680-1080nm) femtosecond Ti:Sapphire oscillator (140fs, 80MHz, Chameleon Ultra-II, Coherent) is used for excitation of CARS-PM, MPEF, and MHG signals. For CARS imaging, a femtosecond pulse from the laser beam is split into a pump/probe beam and another beam synchronously pumping a highly nonlinear polarization maintaining photonic crystal fiber (PCF, FemtoWHITE-800, NKT photonics); the output of the PCF (marked "A" in Fig.1) is used as a synchronized broadband Stokes pulse. A Faraday isolator protects the Ti:Sapphire laser from the back-reflection of the PCF. Laser line filters (i.e., LL01-780, Semrock) are used to reduce the spectral bandwidth of pump/probe pulse. The pump/probe and Stokes pulses (marked "B" and "C" in Fig.1, respectively) are recombined spatially at a long pass filter (such as BLP01-785R, Semrock) and temporally by using delay lines in each beam path and then introduced into a laser scanning unit (FV-300, Olympus). Power and polarization of pulses in different parts of the setup are controlled by half wave plates and Glan laser polarizers. Both pump/probe and Stokes pulses are focused into a sample using an oil-immersion objective (100x, NA=1.4) of an inverted microscope (IX-81, Olympus). A galvano-mirror scans the sample laterally in the focal plane of the objective while the motion of the objective along the microscope's optical axis is controlled by a stepper motor. MNOPM signals are collected by either the same objective (epi-detection mode) and/or another oil-immersion objective (60x, NA=1.42) in forward mode and detected by photomultiplier tubes (H5784-20, Hamamatsu). A series of long-pass dichroic mirrors (i.e., FF735-Di01-25x36, Semrock) and short-pass and band pass filters (BPFs) are used for spectral selection of CARS, MPEF and SHG in the detection channels. We utilized excitation pulses with collinear polarizations controlled by a twisted nematic polarization rotator. Polarization of the forward-detected signals was set by rotating polarizers.

LC samples were prepared between two glass plates of thickness 0.17mm, separated by a gap varied within 10-40μm using spherical particle spacers. To set the surface boundary conditions, we treated their inner surfaces with dilute (2wt.%) aqueous solutions of DMOAP (N,N-dimethyl-n-octadecyl-3-aminopropyl-trimethoxysilyl chloride) for $\mathbf{n(r)}$ perpendicular to the glass plates or with unidirectionally rubbed polyimide (PI-2555, HD Microsystems) coatings for planar alignment. The cells were filled with one of the following room-temperature LCs: ferroelectric SmC* (Felix 015-100), or SmA (8CB, 4-cyano-4-octylbiphenyl), or cholesteric obtained by mixing 5CB (4-cyano-4-pentylbiphenyl) with chiral dopant (cholesteryl pelargonate, Sigma-Aldrich). All LCs were obtained from EM Chemicals. For CARS-PM imaging, we chose the CN stretching vibration (2236 cm$^{-1}$) of 8CB and 5CB molecules parallel to the long molecular axis, since its spectral location is far from other vibrations.

Figure 2 shows three-photon excitation fluorescence (3PF) and SHG forward-detection images of SmC* in an untreated cell obtained using a 1050nm excitation pulse (~1mW) for two orthogonal polarizations. The spectra and selecting filters corresponding to 3PF and SHG images are shown in Fig. 2(i,j). The in-plane images and vertical cross-sections match well for all polarizations of excitation beams (Fig. 2). The strong SHG signal at 525nm [Fig. 2(i)] reveals the polar ordering and corresponding biaxial director structure of the SmC* phase with focal conic domains [1], matching that revealed by 3PF images in Fig. 2(a-h).

To demonstrate the feasibility of simultaneous CARS-PM, 2PF, and 3PF imaging, we used 3µm melamine resin spheres labeled with fluorescein isothiocyanate (FITC) and suspended in 8CB having SmA layers perpendicular to the glass plates of the cell, Fig. 3. The spectra of MNOPM signals along with selection filters corresponding to different imaging modalities are shown in Fig. 3(b,c). Figure 3(d-l) shows in-plane and vertical cross-sectional images of the 30µm-thick sample obtained in three different MNOPM modalities and for excitation light polarization orthogonally to the rubbing direction: (i) 3PF (via self-fluorescence of 8CB due to three photon excitation at 870nm), (ii) CARS-PM (excitation using 780nm pump/probe and broadband Stokes pulses), and (iii) 2PF (fluorescence from FITC-labeled particles with two-photon excitation at 980nm). CARS-PM images due to CN-vibration with the signal centered at ~664nm (CARS frequency is related to that of pump/probe and Stokes pulses as $\nu_{CARS} = 2\nu_{pump/probe} - \nu_{Stokes}$) were forward-detected, Fig. 3(g-i). All images reveal that $\mathbf{n(r)}$ is along the rubbing direction far from the inclusion but distorted around the spherical particle as schematically shown in Fig. 3a. The broadband CARS spectrum of 8CB, Fig. 3(c), shows the capability of imaging by use of other spectral lines, i.e., those due to $\nu(CC)$ and $\nu(CH)$ vibrations.

Submicron resolution of MNOPM along the optical axis is enabled by the nonlinear optical nature of the used processes and is demonstrated using vertical cross-sectional images of ~30μm-thick cells (Fig. 4) that have planar ground-state cholesteric structure shown in Fig. 4(b). 2PF image in Fig. 4(a) shows such a structure with defects and was obtained for a cholesteric LC doped with BTBP (n,n'-bis(2,5-di-tert-butylphenyl)-3,4,9,10-perylene dicarboximide) by the use of 980nm excitation and detection marked on the respective spectrum shown in Fig. 4(c). Similar co-located high-resolution images were obtained in 3PF and CARS-PM modalities without the use of dyes, Fig. 4(d,e).

Images in different modes are consistent with each other and with comparative FCPM studies of similar LC samples. The intensity of detected MNOPM signals depends on the angle between the collinear polarizations of excitation pulses and $\mathbf{n(r)}$ as $\sim \cos^{2i}\theta$ for the detection with no polarizers and as $\sim \cos^{2(i+1)}\theta$ with a polarizer in the detection channel collinear with the polarizations of excitation beams, where $i$ is the order of the nonlinear process. Imaging in 2PF and SHG modes involves second-order nonlinear processes, whereas CARS-PM, 3PF, and third harmonic generation are third-order nonlinear processes. Therefore, MNOPM images in all modalities have a stronger sensitivity to spatial variations of $\mathbf{n(r)}$ compared to single-photon FCPM imaging. Because of the near-infrared excitation, light scattering due to the thermal fluctuations of **n(r)** is relatively small and MNOPM imaging can be done for thick LC sample of thickness ~100μm, which is impossible to achieve using FCPM [2,12]. Compared to the FCPM with the visible-light excitation MNOPM imaging is less affected by artifacts such as those due to light defocusing caused by LC birefringence and the Mauguin effect resulting in the light polarization following the slowly-twisting **n(r)** when the twist occurs in the direction along the microscope's optical axis. We note that FCPM would be unable to visualize the structure of 5CB-based cholesteric shown in Fig. 4 due to the Mauguin effect [12]. The technique can potentially be extended to probe dynamic processes in LCs and LC composites associated with temporal changes of **n(r)** due to application of fields and flow, similar to that recently demonstrated by using the 2PF mode of nonlinear optical microscopy and dye-doped LCs [13]. Additional studies are needed to explore the possible influence of MNOPM excitation beams of typical power ~1mW on the **n(r)** of studied LC structures in various experimental conditions. [8]

In conclusion, MNOPM is capable of non-invasive 3D labeling-free imaging of LC director fields with simultaneous probing of polarity in their self-assembly, enabling direct characterization of devices and displays. This polarizing imaging technique is a cost-effective merging of broadband CARS-PM with MPEF and MHG microscopies and is especially attractive for the study of materials that might exhibit biaxial nematic and smectic phases by use of different chemical bonds of molecules and consecutive comparative polarization analysis of signals.

This work was supported by the Renewable and Sustainable Energy Initiative, International Institute for Complex Adaptive Matter, and by NSF grants DMR-0820579, DMR-0844115, DMR-0645461, and DMR-0847782. We acknowledge discussions with N. Clark and H. Takezoe.

**Figures**

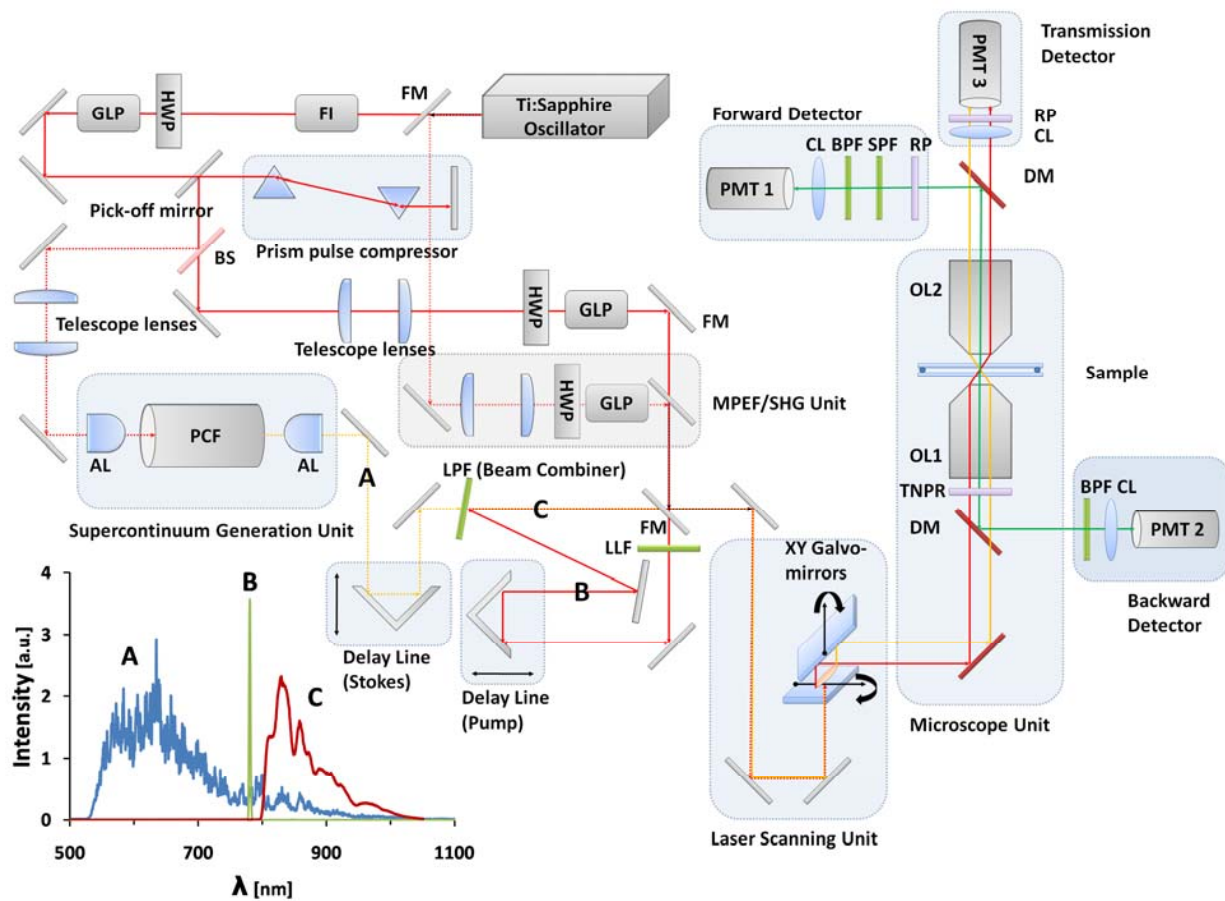

**Fig. 1.** (Color online) Schematic diagram of the MNOPM setup. The inset shows spectra (note that intensity scales are different) at marked positions in the setup: (A) after the PCF, (B) pump/probe pulse at 780nm, and (C) broadband Stokes pulse after the beam combiner. AL: achromatic lenses, BPF: band pass filter, BS: beam splitter, CL: collecting lens, DM: dichroic mirror, FI: Faraday Isolator, FM: flip mirror, GLP: Glan laser polarizer, HWP: broadband half wave plate, LPF: long pass filter, OL: objective lens, PCF: photonic crystal fiber, PMT: photon multiplier tube, RP: rotating polarizer, SPF: short pass filter, TNPR: twisted nematic polarization rotator.

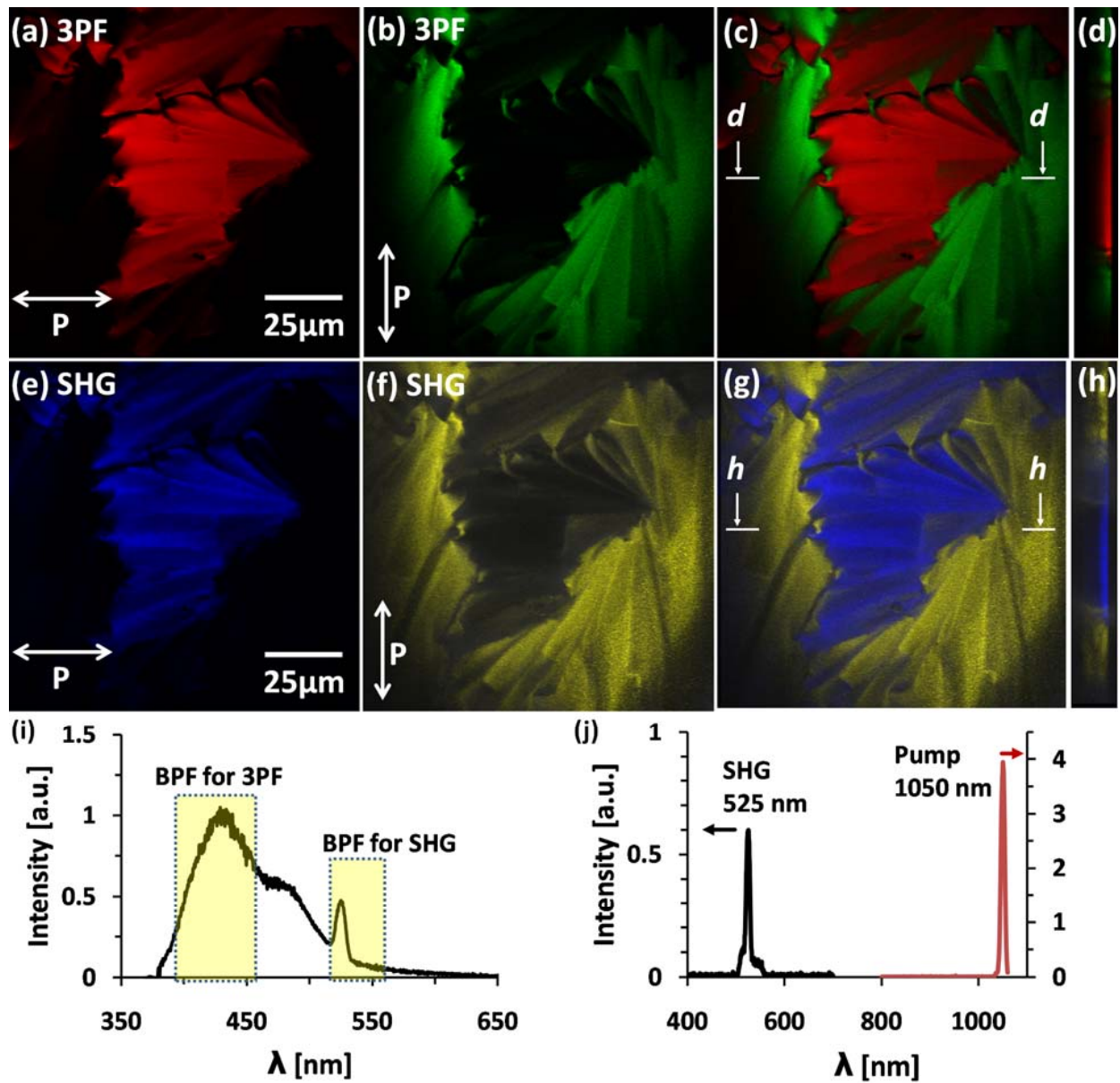

**Fig. 2.** (Color online) Simultaneous 3PF and SHG imaging of a SmC* LC. 3PF images first obtained separately for (a,b) two orthogonal polarizations and then superimposed for (c) in-plane and (d) vertical cross-sections. SHG images first obtained separately for (e,f) two orthogonal polarizations and then superimposed for (g) in-plane and (h) vertical cross-sections. (i) Spectra and filter selections corresponding to the images. (j) Spectra showing the excitation pulse and the generated SHG signal. 3PF and SHG signals were forward-detected using 417/60nm and 535/50nm BPFs, respectively [14].

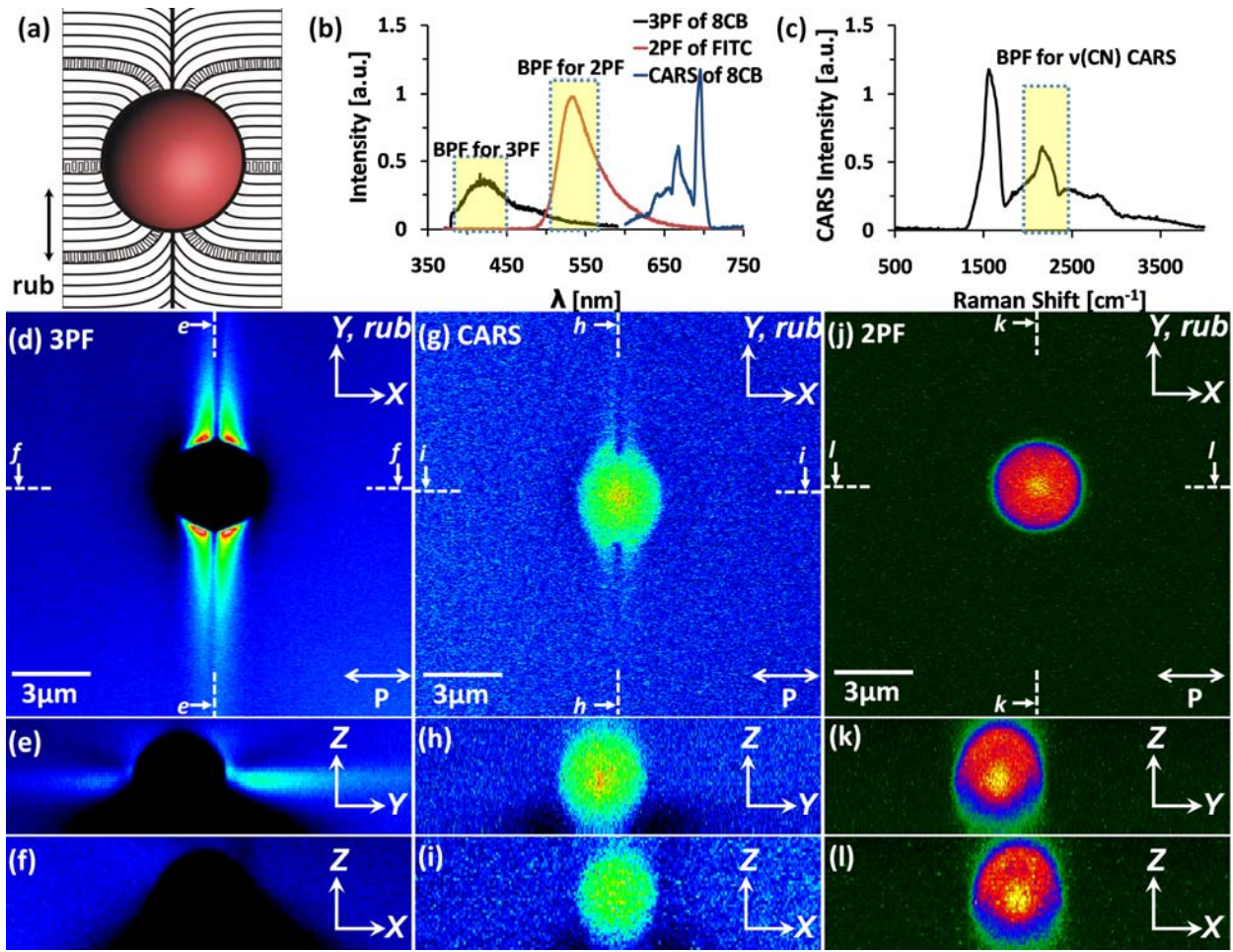

**Fig. 3.** (Color online) MNOPM imaging of LC-colloidal composite. (a) Reconstructed smectic layers and **n(r)** around a particle embedded in an aligned $S_mA$ LC. (b,c) 3PF and broadband CARS spectra of 8CB and 2PF spectrum of FITC (note that the scales are different). 3PF images obtained for excitation at 870nm and detection with a 417/60nm BPF and for (d) XY, (e) YZ, and (f) XZ cross-sections. CARS-PM images obtained using 780nm pump and broadband Stokes pulses for excitation and detection using a 661/20nm BPF and for (g) XY, (h) YZ, and (i) XZ cross-sections. 2PF images of FITC-labeled spheres for excitation at 980nm and detection with a 535/50nm BPF and for (j) XY, (k) YZ, and (l) XZ cross-sections.

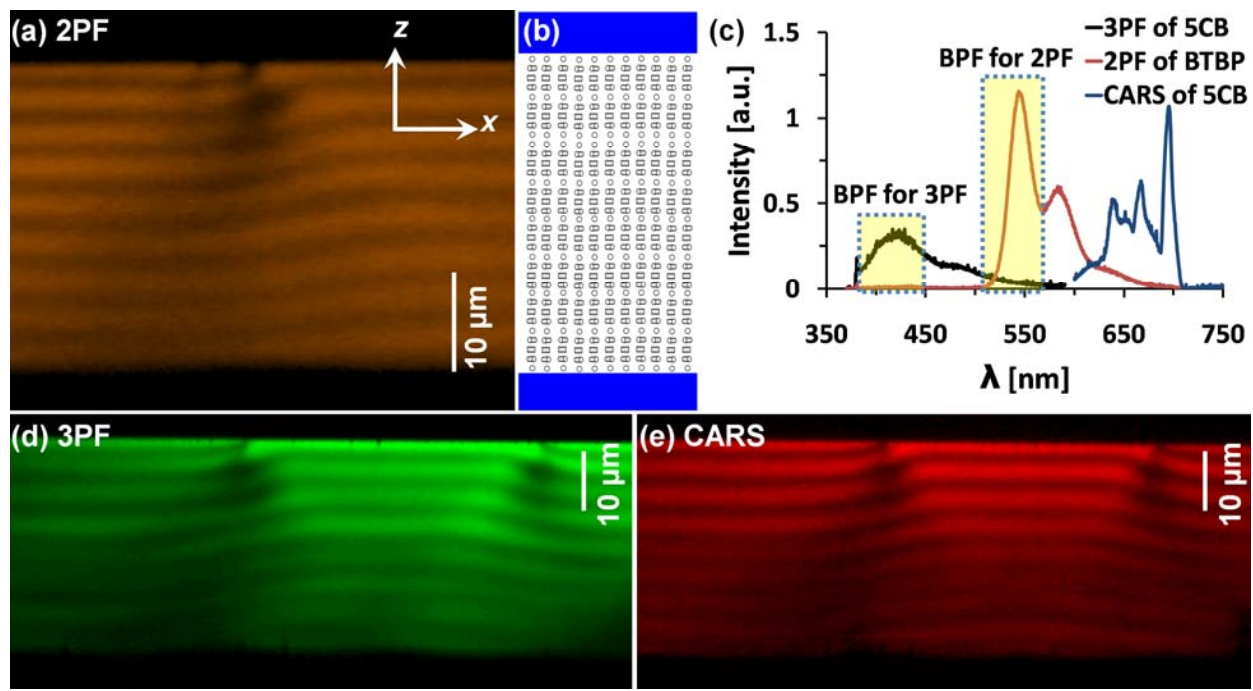

**Fig. 4.** (Color online) Cross-sectional images of a cholesteric LC with 10μm pitch. (a) 2PF image of BTBP doped cholesteric for excitation at 980nm and detection with 535/50nm BPF. (b) Schematics of **n(r)** in a planar cholesteric cell. (c) 2PF spectrum of the BTBP-doped cholesteric LC and 3PF and broadband CARS spectra of an unlabeled cholesteric LC. Images of labeling-free cholesteric LC obtained using (d) 3PF with 870nm excitation and detection with a 417/60nm BPF and (e) CARS-PM with excitation of 780nm pump/probe and a broadband Stokes pulses and detection with a 661/20nm BPF.